\documentclass[final,1p,times]{elsarticle}
\usepackage{hyperref}
\usepackage{graphicx}
\usepackage{amsmath}
\usepackage{amssymb}
\usepackage{mathtools}
\usepackage{color}
\usepackage{bm}
\usepackage[normalem]{ulem}
\usepackage[dvipsnames]{xcolor}
\newcommand*{\FigPath}{./figures}%

\newcommand{\xbj}{\ensuremath{x_{\rm bj}}}

\newcommand{\bt}{b_{\perp}}
\newcommand{\qt}{q_{\perp}}

\newcommand{\eref}[1]{Eq.~(\ref{e.#1})}

\newcommand{\fref}[1]{Fig.~\ref{f.#1}}

\newcommand{\sref}[1]{Sec.~\ref{s.#1}}

\newcounter{bla}

\journal{Computer Physics Communications}

\begin{document}

\begin{frontmatter}

\title{Efficient Fourier Transforms for Transverse Momentum Dependent Distributions}


\author[a,b,c]{Zhong-Bo Kang}
\author[d,e]{Alexei Prokudin}
\author[e,f]{Nobuo Sato}
\author[a,b]{John Terry \corref{author}}

\address[a]{Department of Physics and Astronomy, University of
California, Los Angeles, California 90095, USA}
\address[b]{Mani L. Bhaumik Institute for Theoretical Physics,
University of California, Los Angeles, California 90095, USA}
\address[c]{Center for Frontiers in Nuclear Science, Stony Brook University, Stony Brook, NY 11794, USA}
\address[d]{Division of Science, Penn State University Berks,
Reading, Pennsylvania 19610, USA}
\address[e]{Theory Center, Jefferson Lab, 12000 Jefferson Avenue,
Newport News, Virginia 23606, USA}
\address[f]{Department of Physics, Old Dominion University,
Norfolk, Virginia 23529, USA}

\cortext[author] {John Terry \\\textit{E-mail address:} johndterry@physics.ucla.edu}

\begin{abstract}
Hadron production at low transverse momenta in semi-inclusive deep
inelastic scattering can be described by transverse momentum dependent
(TMD) factorization. 
This formalism has also been widely used to
study the Drell-Yan process and back-to-back hadron pair production in
$e^+e^-$ collisions.
These processes are the main ones for extractions of TMD parton
distribution functions and TMD fragmentation functions, which encode
important information about nucleon structure and hadronization. 
One of the most widely used TMD factorization formalism in
phenomenology formulates TMD observables in coordinate  $b_\perp$-space,
the conjugate space of the transverse momentum. 
The Fourier transform from $b_\perp$-space back into transverse momentum space is sufficiently complicated due to oscillatory integrands that it requires a careful and computationally intensive numerical treatment in order to avoid potentially large numerical errors. 
Within the TMD formalism, the azimuthal angular dependence is analytically integrated and the two-dimensional $b_\perp$ integration reduces to a one-dimensional integration over the magnitude $b_\perp$.
In this paper we develop a fast numerical Hankel
transform algorithm for such a $b_\perp$-integration that improves the numerical
accuracy of TMD calculations in all standard processes. 
Libraries for this algorithm are implemented in Python 2.7 and 3, C++, 
as well as FORTRAN77. All packages are made available open source.
\end{abstract}

\begin{keyword}
CSS formalism, TMDs, Optimized Ogata quadrature
\end{keyword}

\end{frontmatter}

{\bf PROGRAM SUMMARY}

\begin{small}
	\noindent
	{\em Program Title:} Fast Bessel Transform (FBT)                              \\
	{\em CPC Library link to program files:} (to be added by Technical Editor) \\
	{\em Developer's respository link:} https://github.com/UCLA-TMD/Ogata \\
	{\em Code Ocean capsule:} (to be added by Technical Editor)\\
	{\em Licensing provisions:} MIT  \\
	{\em Programming language:} C++, FORTRAN77, Python 2/3\\
	{\em Nature of Problem:} In order to perform extractions of transverse momentum distribution functions, numerical Hankel transforms must be performed from $\bt$-space to momentum space. However, these numerical Hankel transforms are a huge bottleneck in these extractions, making these extractions extremely computationally intensive. \\
	{\em Solution method:}  We develop a numerical Hankel transform algorithm by optimizing Ogata quadrature formula. This algorithm improves the performance of these numerical Hankel transforms by nearly an order of magnitude.\\
\end{small}

\section{Introduction}\label{Introduction}
The transverse momentum dependent (TMD) parton distribution functions
(PDFs) and fragmentation functions (FFs) have received great attention
from both theoretical and experimental communities in recent years.
These TMD PDFs and FFs, or in short called TMDs, provide new
information on hadron structure: the three-dimensional imaging of
hadrons in both longitudinal and transverse momentum
space~\cite{Accardi:2012qut,Boer:2011fh,Aschenauer:2015eha,Perdekamp:2015vwa}.
Significant progress has been made in the last few years in terms of
measuring transverse momentum dependent unpolarized and polarized
cross sections in experiments, as well as extracting the associated
spin-independent and spin-dependent TMDs in phenomenology, see
Refs.~\cite{Boglione:2015zyc, Scimemi:2019gge} and references therein. 

TMDs are non-perturbative objects in Quantum Chromodynamics (QCD) and
thus they have to be either computed on the lattice, or extracted from
experimental data. 
For recent developments on lattice computation  of
TMDs, see Ref.~\cite{Lin:2017snn}.  
On the other hand, in order to extract TMDs from the experimental data, one relies on proper QCD factorization theorems~\cite{Collins:1989gx}. 
TMD factorization~\cite{Collins:2011zzd,Ji:2004wu,Collins:1981uk,GarciaEchevarria:2011rb} describes cross sections in scattering events where the relevant
transverse momentum $q_\perp$ of the observed final state is much
smaller than the hard scale $Q$: $q_\perp\ll Q$.  
In such a region,
the cross section can be factorized in terms of TMD PDFs and/or FFs
and perturbatively calculable short distance hard coefficients.  
The
relevant processes that have been extensively studied include
semi-inclusive deep inelastic scattering
(SIDIS)~\cite{Airapetian:2012ki,Aghasyan:2017ctw}, Drell-Yan type process in proton-proton collisions~\cite{Adamczyk:2015gyk,Aghasyan:2017jop,Aad:2015auj,Aad:2019wmn,Aad:2015lha,Aaboud:2018ezd,Sirunyan:2017igm,Khachatryan:2016nbe,Khachatryan:2016vnn},
and back-to-back dihadron production in $e^+e^-$
collisions~\cite{Leitgab:2013qh,Lees:2013rqd}.
There are also other new opportunities in studying TMDs which are
proposed recently in e.g.
Refs.~\cite{Aad:2011sc,Aidala:2018bjf,Kang:2017glf,Kang:2017btw,Neill:2016vbi,Buffing:2018ggv,Liu:2018trl,Gutierrez-Reyes:2019vbx,Aaij:2019ctd,Seidl:2019jei,Sun:2016kkh,Sun:2014gfa,Cao:2019uor,Sun:2018byn,Sun:2018icb,Chien:2019gyf,Kang:2019ahe,Kang:2020xyq},
and usually involve jet measurements. 

Within the TMD factorization formalism, the cross section is written
as a convolution of the relevant transverse momentum dependent
functions. 
To motivate our discussion and thus make the case more
concrete, let us take SIDIS as an example. 
The differential cross
section for the unpolarized scattering process of $e(\ell)+p(P)\to
e(\ell')+h(P_h)+X$ can be written as~\cite{Bacchetta:2006tn}
\begin{align}
\frac{d\sigma^h}{d\xbj \, dy\,dz\, d^2q_{\perp}}
=\frac{2\pi\alpha^2_{\rm EM}}{Q^2}\frac{1+(1-y)^2}{y}\,
W(q_{\perp},\xbj,z,Q)\,,
\label{e.upol}
\end{align}
where the standard SIDIS variables are defined as
\begin{align}
q=\ell-\ell',\qquad
Q^2=-q^2,\qquad
\xbj=\frac{Q^2}{2P\cdot q},\qquad
y=\frac{P\cdot q}{P\cdot \ell},\qquad 
z=\frac{P\cdot P_h}{P\cdot q}.    
\end{align}
The unpolarized structure functions $W$ in \eref{upol} can be factorized as follows
\begin{align}
W(q_{\perp}, \xbj, z, Q) 
= &\; H(Q,\mu) 
  \sum_q e_q^2 
    \int d^2 {\bf k}_\perp d^2{\bf p}_\perp 
         f_{q/p}(\xbj,k_\perp^2)\,
         D_{h/q}(z, p_{\perp}^2)\,
         \delta^{(2)}\left({\bf k}_\perp + {\bf p}_\perp/z + {\bf q}_{\perp}\right)\,.
\label{eq:sidis_pt}
\end{align}
where ${\bf q}_{\perp}=-{\bf P}_{h\perp}/z$, $e_q$ is the fractional
electric charge for the quarks, and $H(Q, \mu)$ is the hard function
to be given by \eref{HQ} in \sref{Application to TMDs}. On the
other hand, the vectors ${\bf k}_\perp$ and ${\bf p}_\perp$ are the momentum of the produced quark relative to the parent proton and the momentum of the produced hadron with respect to the fragmenting quark, respectively. 
The function $f_{q/p}(\xbj,k_\perp^2)$ is the TMD PDF while $D_{h/q}(z, p_{\perp}^2)$ is the TMD FF. 
Here we have suppressed the additional scale parameters in the TMDs,
which are associated with QCD evolution of the
TMDs~\cite{Kang:2011mr,Echevarria:2012pw,Aybat:2011zv,Aybat:2011ge,Echevarria:2014xaa}.
In general, the convolution and integration of TMDs over the momenta
${\bf k}_\perp$ and ${\bf p}_\perp$ are quite involved. 
Thus in the
original Collins-Soper-Sterman (CSS) approach~\cite{Collins:1984kg},
one takes a Fourier transformation from the momentum space to the
coordinate ${\bf b}_\perp$~space,
~\footnote{There are also other
approaches in the literature that do not work in the ${\bf
b}_\perp$~space, see e.g. Refs.~\cite{Ebert:2016gcn,Kang:2017cjk}. Notice that we drop explicit dependence on $\xbj$ and $z$ for the rest
of this paper.}
\begin{align}
\widetilde{W}(b_\perp, \xbj, z, Q) = 
    \int d^2{\bf q}_{\perp}\; 
    e^{i{\bf q}_{\perp}\cdot {\bf b}_\perp}\; 
   W(q_{\perp}, \xbj, z, Q),
\end{align}
and thus one can write
\begin{align} 
W(q_{\perp}, \xbj, z, Q) 
=& H(Q, \mu)
  \sum_q e_q^2 
    \int \frac{d^2{\bf b}_\perp}{(2\pi)^2}\; 
    e^{-i{\bf q}_{\perp}\cdot {\bf b}_\perp}\; 
    f_{q/p}(\xbj, b_\perp) D_{h/q}(z, b_\perp)\;,
    \nonumber\\
=& H(Q, \mu)
  \sum_q e_q^2\; 
   \int_0^{\infty}\frac{db_\perp b_\perp}{2\pi} J_0(q_{\perp} b_\perp) 
    f_{q/p}(\xbj, b_\perp) D_{h/q}(z, b_\perp)\;,
\label{e.Fuu}
\end{align}
where $b_\perp = |{\bf b}_\perp|$ is the magnitude of the vector ${\bf b}_\perp$, $J_0$ is the Bessel function of the first kind of order~$0$, and $f_{q/p}(\xbj, b_\perp),~D_{h/q}(z, b_\perp)$ are the Fourier transform of the TMD PDF and FF, respectively. In going from the first to second line in~\eref{Fuu}, we perform analytically the integration over the azimuthal angle $\phi$ ,
\begin{align}
    \int_0^{2\pi}\frac{d\phi}{2\pi} e^{-iq_\perp b_\perp \cos(\phi)} =J_0(q_\perp b_\perp)\,,
\end{align}
and thus the two-dimensional Fourier transform reduces to a one-dimensional Hankel transform. 

For the polarized scattering, the generic structure of the cross
sections can be written as~\cite{Ji:2004wu,Bacchetta:2006tn}
\begin{align}
    {\bm q}_\perp^\alpha \,W(q_\perp,\cdots),
    \qquad
    {\bm q}_\perp^\alpha{\bm q}_\perp^\beta \,W(q_\perp, \cdots)
\end{align}
with $W(q_\perp,\cdots)$ representing a generic function of $q_\perp=|{\bm q}_\perp|$ and
``$\cdots$'' denoting the other kinematic variables. 
The Fourier
transform of such functions will lead to Bessel functions of order 1
and 2.  In fact, as shown in~\cite{Boer:2011xd}, all the
spin-dependent structure functions at leading-power can be expressed
in terms of an integration over ${\bm b}_\perp$ multiplied by the
Bessel functions of $J_0$, $J_1$, or $J_2$. Generically the integrals become
\begin{align} 
   \int_0^{\infty}\frac{db_\perp b_\perp^{\nu+1}}{2\pi} J_\nu(q_{\perp} b_\perp) 
    \widetilde{W}(b_\perp)\;,
\label{e.gen}
\end{align}
where the function $\widetilde{W}(b_\perp)$ contains the $b_\perp$ space TMD physics.

Without loss of generality the integration form stemming from the
Fourier transform encountered in TMD observables can be written as
\begin{align}
\int_0^\infty dx \;f(x) \;J_n(x),   
\label{e.general}
\end{align}
where $x = b_\perp \qt$, $J_n(x)$ is the Bessel function of order $n$,
and $f(x)$ is usually a smooth function of $x$ that slowly decays as
$x\to \infty$. 
Such an integration can be extremely computationally intensive and
time consuming with standard integration routines based on adaptive
Gaussian quadratures or Monte Carlo integration methods due to the
oscillatory nature of the Bessel functions.
In the context of TMD global analysis, one has to compute the above
integration many times, and for different kinematic regions, in order
to find the best fit for the non-perturbative TMDs.  
This has become a
huge hurdle for TMD phenomenology in the past for carrying out the
global QCD analysis on TMDs using the data from HERMES, COMPASS, JLAB
6 GeV, Relativistic Heavy Ion Collider (RHIC) and BELLE experiments
and it will become even more challenging for the large amount data
that is expected from the JLab 12 GeV program and the future Electron
Ion Collider (EIC).
Because of this, and because of the complexity of the TMD evolution
improving the efficiency and the speed of the numerical integration of
\eref{general} is extremely desirable.

\begin{figure}[htb!]
\centering
\includegraphics[width=0.5\textwidth]{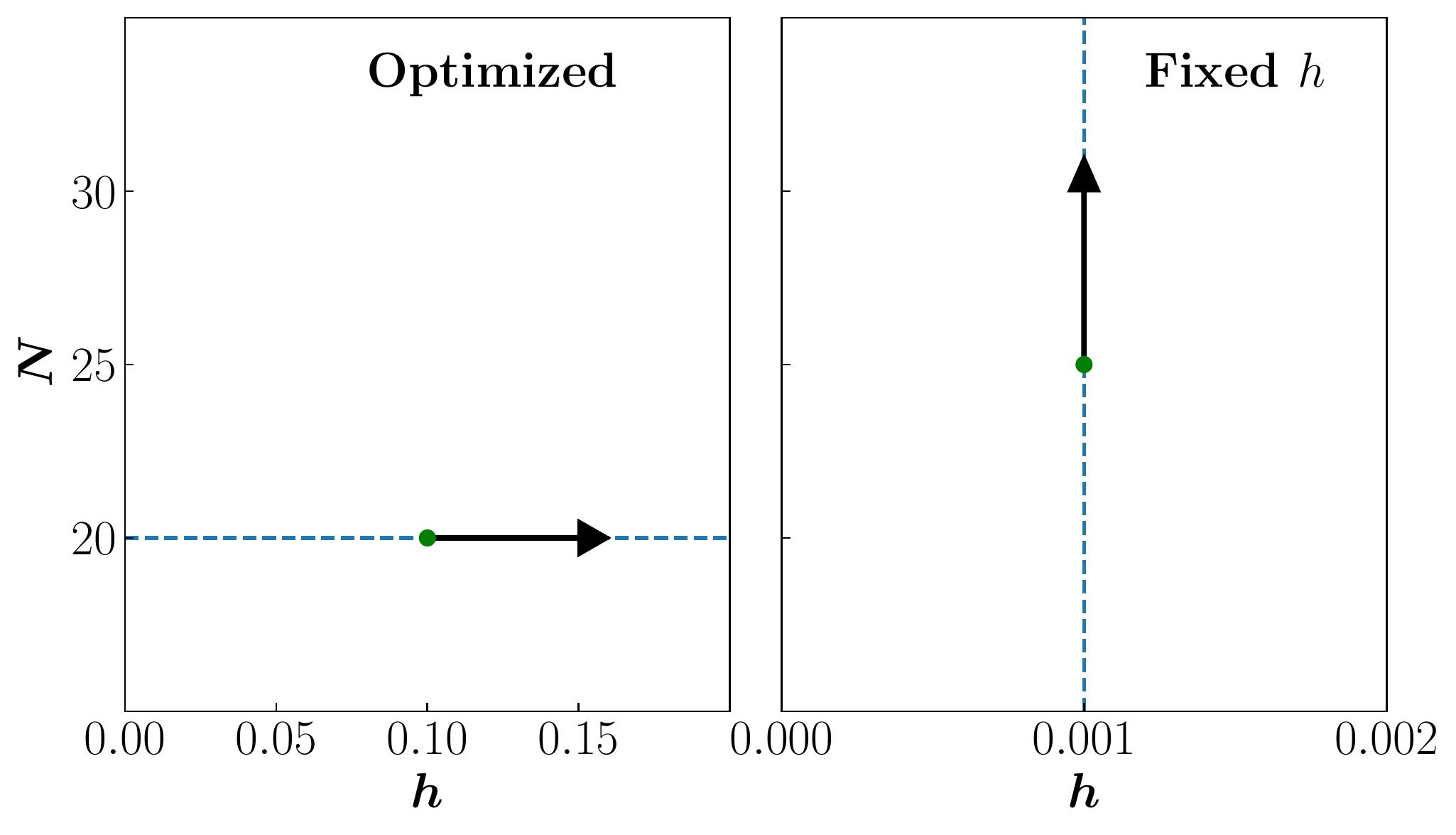}
\caption{Illustration of the schematic difference between the optimized and fixed $h$ Ogata methods. The arrows indicate the flow of the parameters $h$ and $N$ as $\qt$ is increased for each scheme.}
\label{f.cartoon}
\end{figure}

Ogata has introduced a quadrature formula in Ref.~\cite{article} 
that is optimized for integrands with Bessel functions 
for exactly the same integral as in \eref{general}. 
As will be discussed in~\sref{Section2}, this quadrature method has two parameters, $h$ and $N$, which control the 
node spacing and the truncation of the quadrature sum, respectively.
While this formalism has been previously used in TMD analysis in Ref.~\cite{Scimemi:2017etj}; the scheme used in this reference 
suffers from large numerical inefficiencies.
Namely Ref.~\cite{Scimemi:2017etj} method takes the parameter $h$ 
to be held fixed for the entire global analysis, while increasing 
$N$ to reach convergence of the integration. 
We will therefore refer to this method as a `fixed $h$ Ogata scheme'.
In this paper, we propose a new numerical algorithm by which the 
parameter $h$ is optimized for an input number of nodes $N$, 
referred to as `optimized $h$ Ogata scheme'. 
\fref{cartoon} illustrates the difference between these two schemes. 
We note that while the fixed $h$ Ogata method used in Ref.~\cite{Scimemi:2017etj} is sufficient for the extraction of TMDs from several hundred data points, future high precision global extraction of TMDs will utilize tens of thousands data points.
Moreover, the reliable estimation of errors of parameters may warrant Monte Carlo fitting methods, such as ones used in JAM15 extraction from Ref.~\cite{Sato:2016tuz}.

As the Fourier transforms are a huge bottleneck in these extractions, the numerical inefficiencies of fixed $h$ Ogata methods will be massively amplified to the point where precision extractions may no longer be feasible. 
We demonstrate in this paper that the optimized Ogata
quadrature is more efficient than the traditional adaptive Gaussian
quadrature method, Monte Carlo method, as well as fixed $h$ 
Ogata quadrature method.
The rest of this paper is organized as follows.
In \sref{Section2}, we summarize the relevant formalism for the
Ogata quadrature method and describe our optimized numerical algorithm
in detail. 
In \sref{Benchmarking the Numerical Precision}, we give a
demonstration of the optimized Ogata quadrature and benchmark the
algorithm against adaptive Gaussian quadrature using an exponential
function which has an analytic Fourier transform.
In \sref{Application to TMDs}, we apply our numerical method to an
example of a phenomenological form of TMDs. 
We conclude our paper in \sref{conclusions}.

\section{Optimized Ogata Quadrature Formalism}
\label{s.Section2}

In this section we first review the original Ogata quadrature
formalism and then discuss our optimization scheme for performing
high efficiency numerical integrals relevant to TMD analysis.
The Ogata method is based on a quadrature formalism 
that was first introduced in Ref.~\cite{Frappie} by Frappier and Olivier. 
The quadrature formula for the
integrand of the form $|x|^{2n+1}f(x)$
reads:
\begin{align}
\int_{-\infty}^{\infty} dx\, |x|^{2n+1}f(x) 
  = h \sum_{j=-\infty, j\neq 0}^{\infty}  w_{n j}\, |x_{n j}|^{2n+1}\,f(x_{n j}) 
    + {\mathcal O}\left(e^{-c/h}\right)\; ,
\label{e.quad_init}
\end{align}
where the function $f(x)$ must be an integrable function for the sum
to be finite. The nodes $x_{nj}$ and the weights $w_{nj}$ of the
quadrature are given by
\begin{align}
x_{nj} = h \xi_{n j}  \; ,
\qquad
w_{n j} = \frac{2}{\pi^2 \xi_{n |j|}J_{n+1}(\pi\xi_{n |j|})}\;,
\end{align}
with $j=\pm1, \pm2, \cdots$, and $\xi_{nj}$ the zeros of the Bessel
function $J_n(\pi x)$ of order $n$, i.e. $J_{n}(\pi \xi_{n j}) = 0$,
and the parameter $1/h$ represents the node density.  
The term
${\mathcal O}\left(e^{-c/h}\right)$ accounts for the error of the
quadrature sum approximation at a finite $h$ as described in
equation~(2.2) of Ref.~\cite{article}, and $c$ is a positive constant,
whose precise value depends on the functional form of $f(x)$. 
For the
time being, we will be interested in the case of $f(x)$ being an even
function of $x$ which results in the following quadrature formula
\begin{align}
\int_{0}^{\infty} dx \,x^{2n+1}\,f(x) 
  = h \sum_{j=1}^{\infty} w_{n j} \,x_{n j}^{2n+1}\,f(x_{n j}) 
    + {\mathcal O}\left(e^{-c/h}\right).
\label{e.quad}
\end{align}
In practice the sum in \eref{quad} is truncated at a given $j = N$
which introduces an error of 
\begin{align}
    \mathcal{I}_{n \, N+1} = h \sum_{j=N+1}^{\infty} w_{n j} \,x_{n j}^{2n+1}\,f(x_{n j})
\label{e.trun}    
\end{align}
and the quadrature formula  becomes
\begin{align}
 \int_{0}^{\infty} dx \,x^{2n+1}\,f(x) 
  =h \sum_{j=1}^{N} w_{n j} \,x_{n j}^{2n+1}\,f(x_{n j})
   +\left[\mathcal{I}_{n \, N+1} + {\mathcal O}\left(e^{-c/h}\right)\right].
\end{align}

The following change of variables, see Ref.~\cite{article}, optimizes
the convergence of integrals of the typical TMD functional form from
\eref{general}:
\begin{align}
x=\frac{\pi}{h}\psi(t)    
\qquad
\mbox{with \; $\psi(t) = t \tanh\left(\frac{\pi}{2}\sinh t\right)$}\;\;,
\label{e.transform}
\end{align}
so that \eref{general} becomes
\begin{align}
\int_{0}^{\infty} dx\, f(x) J_n(x) 
  =&\frac{\pi}{h} \int_{0}^\infty dt \; 
      \psi'(t)
      f\left(\frac{\pi}{h}\psi(t)\right) 
      J_n\left(\frac{\pi}{h}\psi(t)\right)   
\nonumber\\
  =& \frac{\pi}{h} \int_{0}^\infty dt \;
      |t|^{2n+1} 
      \frac{  \psi'(t)
              f\left(\frac{\pi}{h}\psi(t)\right)
              J_n\left(\frac{\pi}{h}\psi(t)\right)}
      {t^{2n+1}}\;,
\end{align}
where $\psi'(t)=d\psi(t)/dt$.  
At this point, it is important to
realize that the part of the integrand beside the factor $|t|^{2n+1}$
is an even function of $t$, and thus we can apply \eref{quad} and
obtain the following quadrature formula
\begin{align}
\int_{0}^{\infty} dx \,f(x)\,J_n(x) 
  = \pi \sum_{j=1}^{N} 
      w_{n j} \,
      f\left(\frac{\pi}{h}\psi(x_{n j})\right) \,
      J_n\left(\frac{\pi}{h}\psi(x_{n j})\right) \,
      \psi'(x_{nj}) 
+\left[ \mathcal{I}'_{n\, N+1} + {\mathcal O}\left(e^{-c/h}\right) \right]\,,
\label{e.trans}
\end{align}
where $\mathcal{I}'_{n\, N+1}$ are the same truncation errors defined
in \eref{trun} but with the transformed integrand,
\begin{align}
\mathcal{I}'_{n\, N+1} 
  = \pi \sum_{j=N+1}^{\infty} 
      w_{n j} \,
      f\left(\frac{\pi}{h}\psi(x_{n j})\right) \,
      J_n\left(\frac{\pi}{h}\psi(x_{n j})\right) \,
      \psi'(x_{nj})\,.
\label{e.errornew}      
\end{align}
\eref{trans} is the aforementioned Ogata quadrature formula, which we
advocate in our current paper. The variable substitution has the
useful asymptotic behavior
\begin{align}
\frac{\pi}{h}\psi(x_{n j}) 
  \approx \pi \xi_{n j} \left[1-2\exp \left( -\frac{\pi}{2}e^{x_{n j}} \right)\right],
\end{align}
such that the asymptotic behavior for the Bessel function becomes
\begin{align}
 J_n\left(\frac{\pi}{h}\psi(x_{n j})\right)\;  
  \approx 2 \pi \xi_{n j} J_{n+1}(\pi \xi_{n j})
          \exp \left( -\frac{\pi}{2}e^{x_{n j}} \right).
\label{e.assympt}
\end{align}
This variable substitution then enforces the double exponential
convergence of the quadrature sum in $j$.

The quadrature sum has two parameters, $h$ and $N$, which control the
efficiency and the magnitude of the error terms. To generate a high
efficiency algorithm, the numerical integration must be performed with
a small $N$ while at the same time the error terms must also be small,
to ensure reliable results.  By inspecting  Eqs.~\eqref{e.errornew}
and \eqref{e.assympt} one notes that for a small number of function
calls the truncation errors will be large if $h$ is too small. 
To compensate for this, the numerical algorithm would need to sum 
a large number of nodes $N$ to minimize truncation errors. Small 
values of $h$ will then tend to generate numerical inefficiencies. This
has been the leading cause of numerical inefficiencies in previous
implementations of the Ogata quadrature method.
At the same time for larger values of $h$ the quadrature error grows as
$\sim e^{-c/h}$, see \eref{quad_init}. These observations indicate
the need to find optimal values for $h$ and $N$ that keep the error
term in \eref{trans} as small as possible. 
We found that such optimal values can be obtained by enforcing the
largest contribution to the quadrature to be the first term in the
truncated sum of \eref{quad} which can be achieved by maximizing the
contribution of the first node, i.e.
\begin{align}
\frac{\partial}{\partial h}\left( h (h \xi_{n 1})^{2n+1} f(h \xi_{n 1}) \right) 
= 0\,.
\label{e.opt}
\end{align}
By solving numerically \eref{opt} for $h$ one finds the optimal value
of $h$ for the quadrature method in \eref{quad}. We will refer to this
optimal value  as $h_{\rm u}$. 

It is now worth noting that $h_{\rm u}$ will tend to be a large value. 
This makes intuitive sense since minimizing truncation errors can be 
achieved by using a large spacing parameter. However, taking a large value of $h$ 
introduces quadrature errors which behave like $e^{-c/h}$ and tend to be large 
for $h = h_{\rm u}$. 
This issues can be mitigated by using the following scheme. 
We first use the condition in \eref{opt} to minimize truncation errors
in \eref{quad}.
We then impose the condition that the final nodes of Eqs.~\eqref{e.quad} and \eqref{e.trans}
are placed at the same location by enforcing that
\begin{align}
h_{\rm u} \xi_{n N} = \frac{\pi}{h} \psi(h \xi_{n N})\;.
\label{e.opt_t}
\end{align}
This ensures that the quadrature in \eref{trans} has the same
truncation errors as \eref{quad} with $h_{\rm u}$. The solution for
$h$ in the above equality is  given by
\begin{align}
h=\frac{1}{\xi_{n N}}\sinh^{-1}\left( \frac{2}{\pi} \tanh^{-1}\left(\frac{h_{\rm u}}{\pi}
\right) \right)\,,
\label{e.huht}
\end{align}
when $h_u<\pi$.
This value, labeled as $h_t$ is the optimal value for $h$ to
be used in \eref{trans}.
Note that  $h_{\rm t}$ is suppressed by a large factor of $\pi\xi_{n N}$
so that $h_{\rm t}\ll h_{\rm u}$.
In \fref{ht} we plot the ratio $h_{\rm t}/h_{\rm u}$ as a
function of $h_{\rm u}$ for $N = 10,20$, and $40$. We find that in all
cases $\frac{h_{\rm t}}{h_{\rm u}}\ll 1$ which avoids large errors 
in \eref{opt}. 
We note that \eref{opt_t} only has a real solution when $h_u<\pi$. It turns out that when the variance of the input function is very large, the value of the parameter $h_u$ which was determined from \eref{opt} may be larger than $\pi$. To ensure that there is a real solution to \eref{opt_t} as well as to ensure that $h_t<h_u$ for all values of $N$, we must set an upper boundary on the parameter $h_u$, which we set to be $h_{max} = 2$. When this occurs, the value of $h_u$ that we use is smaller than the optimal value. This issue could lead to truncation errors if the variance of the input function is sufficiently large. Beyond the range, the user needs to rely on the number of sampling points $N$ in order to decrease the errors stemming from the truncation errors. 

\begin{figure}[htb!]
\centering
\includegraphics[width=0.5\textwidth]{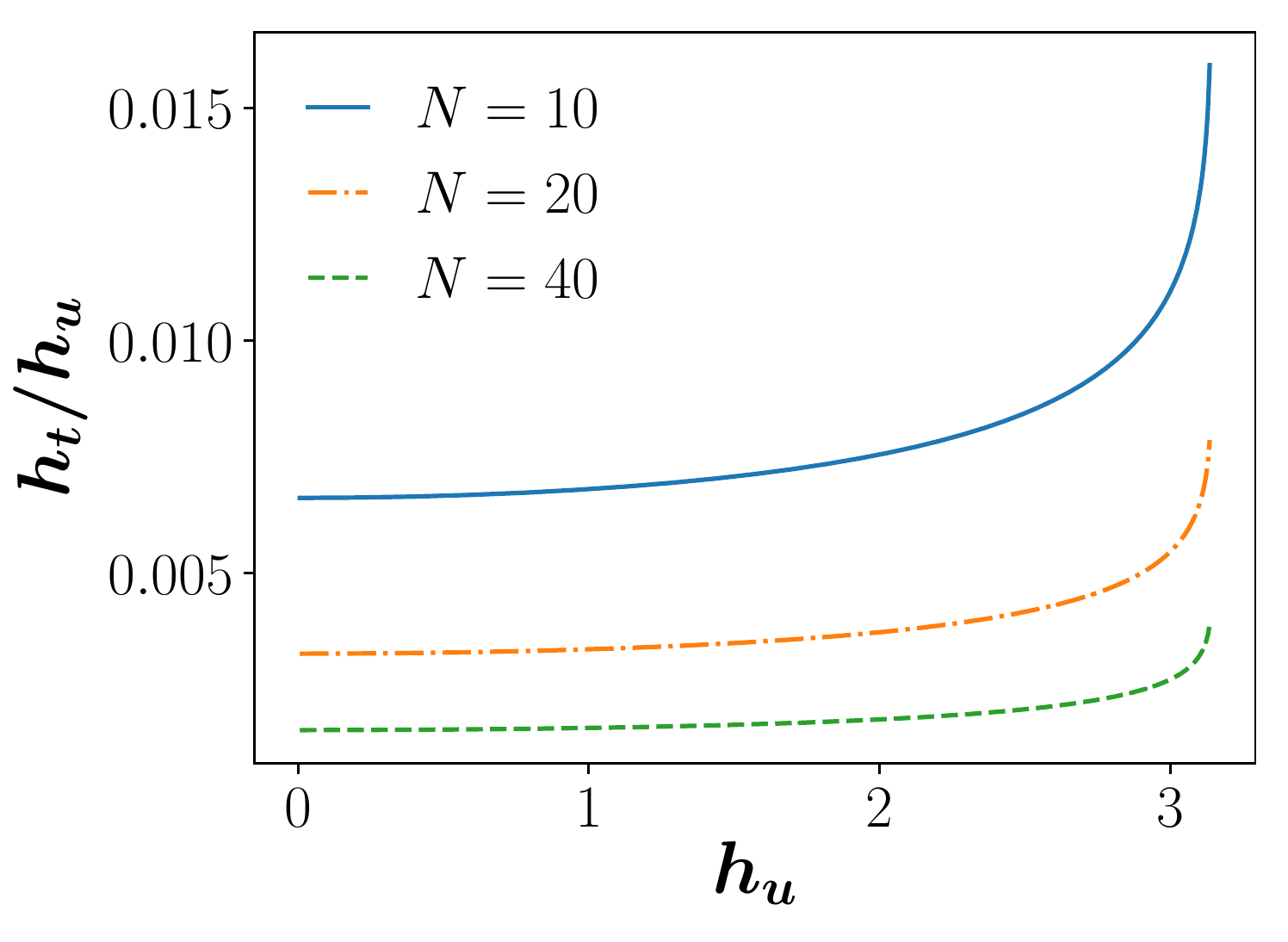}
\caption{The solution of \eref{huht} as a function of $h_{\rm u}$ at several
  values of $N$. The solution is written as $h_{\rm t}/h_{\rm u}$ to demonstrate
  that $h_{\rm t}\ll h_{\rm u}$ for $h_{\rm u}<\pi$.}
\label{f.ht}
\end{figure}

In summary, for a given choice of number of integrand evaluations $N$,
our procedure minimizes the error contribution in $h$ as well as
truncation errors by applying the conditions Eqs.~\eqref{e.opt} and~\eqref{e.opt_t}.  The application of these conditions determines an optimal
choice for $h$ in implementing the quadrature formula of \eref{trans}.
We shall refer to this procedure as ``the optimized Ogata quadrature
formula.''  We will demonstrate below the efficiency of our formalism,
first through the use of toy TMDs, and then through QCD based TMDs.

\section{Benchmarking the Numerical Precision}
\label{s.Benchmarking the Numerical Precision}

In this section, we demonstrate the efficiency of the optimized Ogata
quadrature method using toy TMDs for which the exact Fourier-Bessel transform
is known.  We will compare the numerical efficiency of the optimized
Ogata quadrature against adaptive Gaussian quadrature, which is
available in QUADPACK integration routine in
Ref.~\cite{1983qspa.book.....P}. 
It is important to emphasize that even though we mainly demonstrate
the method for the integration involving Bessel function $J_0(x)$, we
have checked that it works equally well for integration involving
either $J_1(x)$ or $J_2(x)$, relevant for TMD studies in polarized
scattering.

To assess the efficacy of our quadrature method we will map the
error of the integration relative to the exact known result as a
function of number of integrand calls. As discussed before we are
interested in performing integrals of the form
\begin{align}
W(\qt) = \int_0^{\infty} \frac{d \bt \bt}{2\pi}\; \widetilde{W}(\bt)\;
J_{0}(\bt \qt) \; ,
\label{e.hankel}
\end{align}
where  the function $\widetilde{W}(\bt)$ contains the $\bt$ space TMD
physics. Therefore we will use a toy $\widetilde{W}(\bt)$ which mimics
the ${\bm b}_\perp$ space behavior of realistic unpolarized TMDs that
has an exact analytic Fourier-Bessel transform. 
Specifically we choose the gamma distributions which are given in
terms of the distribution's mean, $\beta$, and variance, $\sigma$, as
\begin{align}
\widetilde{W}(\bt,\beta,\sigma) 
    = \frac{1}{\bt}\left(\frac{\beta \bt}{\sigma^2}\right)^{\beta^2/\sigma^2}
      \frac{e^{
            -\frac{\beta\;\bt}{\sigma^2}}}
              {\Gamma\left(\frac{\beta^2}{\sigma^2}\right)}\;.
\label{e.wtildetoy}
\end{align}
This function has an exponential $b_\perp$-dependence, and has been
used in the literature for TMD studies~\cite{Bertone:2019nxa}.  Its
exact Fourier-Bessel transform is given by
\begin{align}
W^{\rm exact}(\qt,\beta,\sigma)  = \frac{1}{2\pi}\left(\frac{\sigma
   ^{2}}{\beta}\right)
      \frac{\Gamma \left(\frac{\beta
   ^2}{\sigma ^2}+1\right)
   }{\Gamma \left(\frac{\beta ^2}{\sigma
   ^2}\right)} \,_2\tilde{F}_1\left(\frac{1}{2} \left(\frac{\beta
   ^2}{\sigma ^2}+1\right),\frac{1}{2}
   \left(\frac{\beta ^2}{\sigma ^2}+2\right);1;-\frac{\qt^2 \sigma ^4}{\beta
   ^2}\right)\;,
   \label{e.exact}
\end{align}
where $_2\tilde{F}_1\left(a,b;c;d\right)$ is the regularized Gaussian
hyper-geometric function.
\begin{figure}[htb!]
\centering
\includegraphics[width=\textwidth]{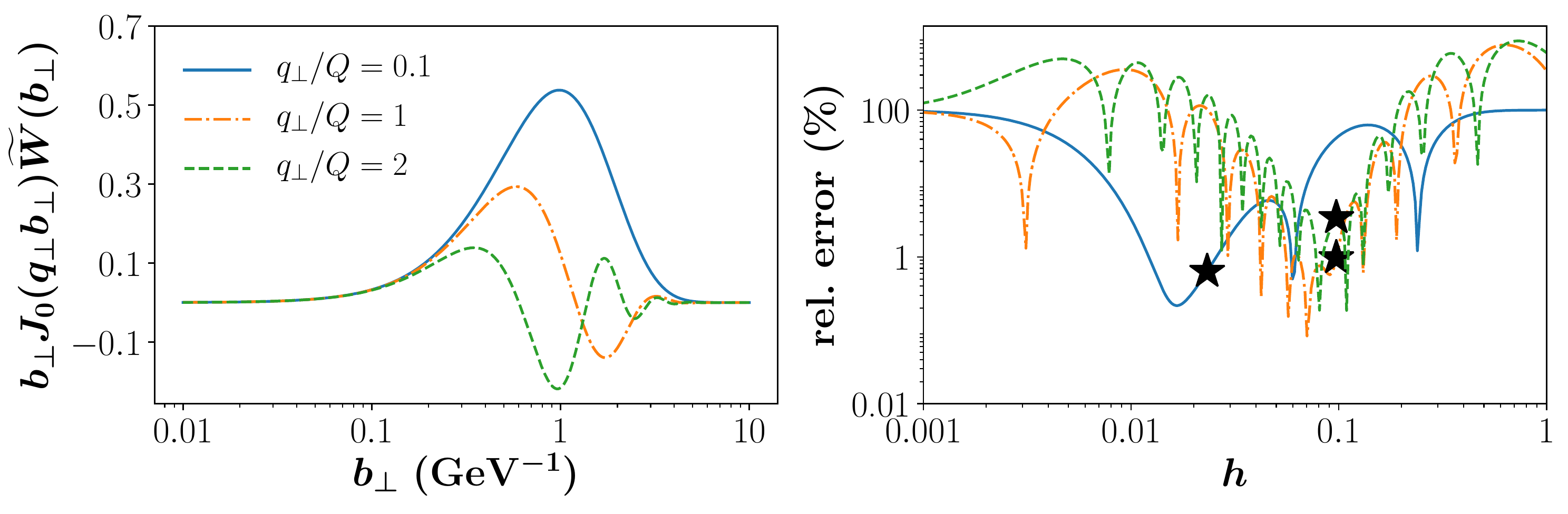}
\caption{{\bf Left panel:} The integrand of Eq.~\eqref{e.hankel} with
  $\widetilde{W}$ from Eq.~\eqref{e.wtildetoy} as a function of $\bt$
  for small, moderate and large transverse momenta $\qt = 0.2,~2$, and
  $4$ GeV. We choose $Q=2$ GeV in \eref{QQ}. {\bf Right panel:} The
  relative percent error \eref{error} of the Ogata quadrature
  is plotted as a function of $h$ for small, moderate and large
  transverse momentum $\qt$ with $N = 4,7,10$, the number of nodes used in the
  integration. The stars indicate the value of $h$ which is
  determined from the optimization conditions Eqs.~\eqref{e.opt} and \eqref{e.opt_t}.}
\label{f.Wtest}
\end{figure}
The function $\widetilde{W}(\bt,\beta,\sigma)$ has a single peak in
$\bt$ space, which is given in terms of $\beta$ and $\sigma$ as 
\begin{align}
\bt^{\rm peak} = \frac{\beta^2-\sigma^2}{\beta}.
\end{align}
We further introduce a parameter $Q$ to our toy TMD model, which is the
inverse of the $\bt^{\rm peak}$, i.e.
\begin{align}
Q = \frac{1}{\bt^{\rm peak}} = \frac{\beta}{\beta^2-\sigma^2}.  
\label{e.QQ}
\end{align}
Such a $Q$-dependence mimics the hard scale $Q$ encountered in QCD
based TMDs such as the photon virtuality in SIDIS reactions
~\cite{Collins:1984kg,Qiu:2000hf}. 
Notice that it is the quantity  $q_\perp/Q$ that controls how
oscillating the toy TMD is. The larger $q_\perp/Q$ is, the more
oscillating the integrand is in $\bt$ space and the more numerically
intensive the integration becomes. 

For our demonstration, we take $Q = 2$ (GeV) and $\sigma = 1$ (GeV$^{-1}$)
similar to the usual JLab kinematics. We choose $\qt = 0.2, 2$, and
$4$ (GeV), and plot the integrands on the left hand side of
\fref{Wtest}. As one can see clearly, the integrands do become more
oscillating as $q_\perp/Q$ increases. To test the precision of our
formalism, we take as an example in our optimized Ogata formula in
\eref{trans}. 
\begin{align}
 W^{\rm Ogata}(\qt,\beta,\sigma) =   \pi \sum_{j=1}^{N} w_{n j} \,f\left(\frac{\pi}{h}\psi(h \xi_{n j})\right) \,J_n\left(\frac{\pi}{h}\psi(h \xi_{n j})\right) \,\psi'(h \xi_{nj})
\end{align}
The relative percent error is defined as 
\begin{align}
{\rm rel.\; error}\;(\%) =
\left|
\frac{W^{\rm exact}(\qt,\beta,\sigma) - W^{\rm Ogata}(\qt,\beta,\sigma)}
      {W^{\rm exact}(\qt,\beta,\sigma)}
\right|
\times 100\;,
\label{e.error}
\end{align}
where the exact result $W^{\rm exact}(\qt,\beta,\sigma)$ is given in
\eref{exact}. 
For this analysis, we increase the number of nodes at $\qt = 0.2, 2, 4$ GeV until the best relative error for the numerical inversion is of the order of one percent. This requires $4,7$ 
and $10$ nodes at $\qt = 0.2, 2, 4$ GeV, respectively. On the right panel of 
\fref{Wtest}, we plot the relative
percent error of the numerical integration as a function of the
parameter $h$ for $\qt = 0.2$, $2$, and $4$ GeV. One can
see that in each case, there is an optimal value of the parameter $h$,
which minimizes the measured error. 

Intuitively having a small node spacing $h$ should result in a small
error, since the error in $h$ is of the order $\mathcal{O}(e^{-c/h})$
in \eref{trans}. However, the truncation errors 
$\mathcal{I}'_{n\,N+1}$ in \eref{trans} will generate large errors in the numerical
integration, due to the factors of $\,f\left(\frac{\pi}{h}\psi(h
\xi_{n j})\right) \,J_n\left(\frac{\pi}{h}\psi(h \xi_{n j})\right)$,
unless one increases $N$ significantly. 
Therefore, for a small and fixed number of nodes $N$, there is an
optimized $h$ that minimizes the errors as argued in \sref{Section2}. 
On the right panel of \fref{Wtest}, we indicate with stars the values
of $h$ which are determined by the optimization conditions
Eqs.~\eqref{e.opt} and \eqref{e.opt_t}. We find that within this range
of kinematics, our optimization conditions indeed determine suitable
values of $h$ for our quadrature method which is key to achieve high
efficiency in the numerical integration.
\begin{figure}[htb!]
\centering
\includegraphics[width=\textwidth]{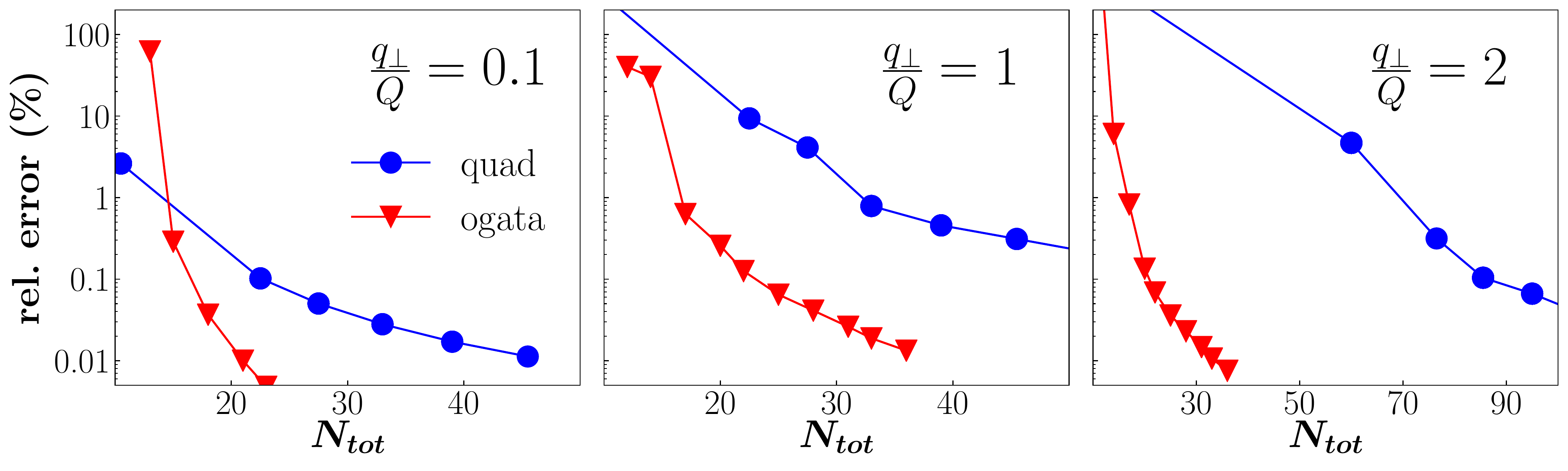}
\caption{From left to right, the relative percent error  of optimized
  Ogata and adaptive Gaussian quadrature as a function of total number
  of calls $N_{tot}$ to the integrand at small, moderator and large
  transverse momentum: $\qt/Q=0.1$ (left), $\qt/Q=1$ (middle), and
  $\qt/Q=2$ (right).}
\label{f.Wtestn}
\end{figure}

We now compare the efficiency of the optimized Ogata quadrature
against adaptive Gaussian quadrature.  Note that the optimized Ogata
quadrature first samples the integrand to determine the optimal value
of $h$ using \eref{opt}. The integration routine then samples the
integrand at $N$ nodes to perform the sum. This results in a total
number of integrand calls $N_{tot}$. To control $N_{tot}$
for adaptive Gaussian quadrature, we change the relative error tolerance of the integrator.
Likewise, we measure the total
number of function calls of adaptive Gaussian quadrature requested by
the numerical routine.
In \fref{Wtestn} we plot the relative error as a function of
$N_{tot}$ for small, intermediate and large values of $q_{\perp}/Q$.
As is evident, the optimized Ogata quadrature method is more
efficient than adaptive Gaussian quadrature, for all three probed
regions of $q_\perp/Q$ with relative errors that go below $0.1\%$.

\section{Application to TMDs}
\label{s.Application to TMDs}

In this section we use the optimized Ogata quadrature to calculate the
SIDIS $q_\perp$-differential cross sections in QCD TMD factorization
framework. We then use these calculations to describe COMPASS charged
hadron multiplicity data~\cite{Aghasyan:2017ctw}. In addition, we use
adaptive Gaussian quadrature and Vegas Monte Carlo algorithm for the
same calculations to benchmark the performance.

Let's first summarize the basic ingredients for the  implementation of the unpolarized SIDIS structure function $W$ in~\eref{Fuu} in the CSS TMD framework~\cite{Collins:2011zzd,Aybat:2011zv}.  In
such context the TMD PDFs and TMD FFs can be expressed as
\begin{align}
f_{q/p}(\xbj, b_\perp, \mu, \zeta) 
    = \sum_j \int_{\xbj}^1 & \frac{d\hat{x}}{\hat{x}} 
      C_{q/j}^{pdf}\left (\xbj/\hat{x},b_{*},\mu_{b_*}\right )
      f_{j/p}\left (\hat{x},\mu_{b*}\right ) 
    \\
    & \times \exp \left( S_{\rm pert}-g_A(\xbj,\bt,b_{max})
                 -\frac{1}{2}g_K(\bt,b_{max})
                 \ln\left (\frac{\sqrt{\zeta}}{Q_0}\right ) 
           \right)\;, \nonumber
\\
D_{h/q}(z, b_\perp,  \mu,\zeta) 
  = \sum_j & \int_{z}^1\frac{d\hat{z}}{\hat{z}}
     C_{j/q}^{ff}\left (z/\hat{z},b_{*},\mu_{b_*}\right)
      d_{h/j}\left(\hat{z},\mu_{b*}\right ) \\
    & \times   \exp \left(S_{\rm pert}-g_B(z,\bt,b_{max})
                -\frac{1}{2}g_K(\bt,b_{max})
                 \ln\left (\frac{\sqrt{\zeta}}{Q_0}\right ) 
          \right)\;, \nonumber
\end{align}
where $\mu$ is the renormalization scale, $\zeta$ is the rapidity
scale, $C_{q/j}^{pdf}$ and $C_{j/q}^{ff}$ are perturbatively
calculable coefficient functions (see Ref.~\cite{Aybat:2011zv}), and
$f_{j/p}\left (\hat{x},\mu_{b*}\right )$ and
$d_{h/j}\left(\hat{z},\mu_{b*}\right )$ are the standard collinear
PDFs and FFs, respectively.  We will use the initial scale $Q_0^2 =
2.4$ GeV$^2$.
We follow the usual $b_*$-prescription~\cite{Collins:2011zzd} to avoid
the Landau pole of $\alpha_s$, with
\begin{align}
b_* = \frac{\bt}{\sqrt{1+\bt^2/b_{max}^2}}. 
\end{align}
The perturbative Sudakov factor $S_{\rm pert}$ is given by
\begin{align}
S_{\rm pert} = 
  \frac{1}{2} \int_{\mu_{b_*}}^{\mu}\frac{d\mu'}{\mu'}
      \left[2\gamma(\mu')-\ln\left({\frac{\zeta}{{\mu'}^2}}\right)\gamma_K(\mu')\right]
      +\widetilde{K}(\bt,\mu_{b_*})\ln\left (\frac{\sqrt{\zeta}}{\mu_{b_*}}\right ),
\end{align}
i.e., the evolution is done from the auxiliary scale $\mu_{b_*} =
2e^{-\gamma_E}/b_{*}$ to the  scale $\mu$.  In the actual
phenomenology, we set the rapidity scale $\zeta=Q^2$ and set the
renormalization scale $\mu=Q$.
We will implement the TMD evolution at next-to-leading-logarithmic
(NLL) accuracy, and use the coefficient functions $C$ at
next-to-leading order (NLO). All the relevant NLO coefficients and NLL
anomalous dimensions can be found in
Refs.~\cite{Collins:2011zzd,Aybat:2011zv}. In addition, we use NLO
expression for hard function $H(Q,\mu)$ in \eref{Fuu} from
Ref.~\cite{Aybat:2011zv} reads
\begin{align}
H(Q, \mu) = 1+\frac{\alpha_s}{\pi}C_F 
      \left[ \frac{3}{2}\ln\left(\frac{Q^2}{\mu^2}\right) 
            -\frac{1}{2}\ln^2\left( \frac{Q^2}{\mu^2}\right) -4 \right]\, ,
\label{e.HQ}
\end{align}
and we set  $\mu=Q$  so that the logarithmic terms vanish.  Finally,
we choose the parametrizations for the non-perturbative factors used
in Refs.  ~\cite{Kang:2015msa,Su:2014wpa} which are given by
\begin{align}
g_A(\xbj,\bt,b_{max}) = g_q \bt^2,\,
\qquad
g_B(z,\bt,b_{max}) = \frac{g_h}{z^2}\bt^2\,,
\qquad
g_K(\bt,b_{max}) = g_2\ln\left(\frac{\bt}{b_*}\right)\,,
\end{align}
with  $g_q = 0.106$ GeV$^2$, $g_2 = 0.84$, and $g_h = 0.042$ GeV$^2$. 
\begin{figure}[t!]
\centering
\includegraphics[width=0.5\textwidth]{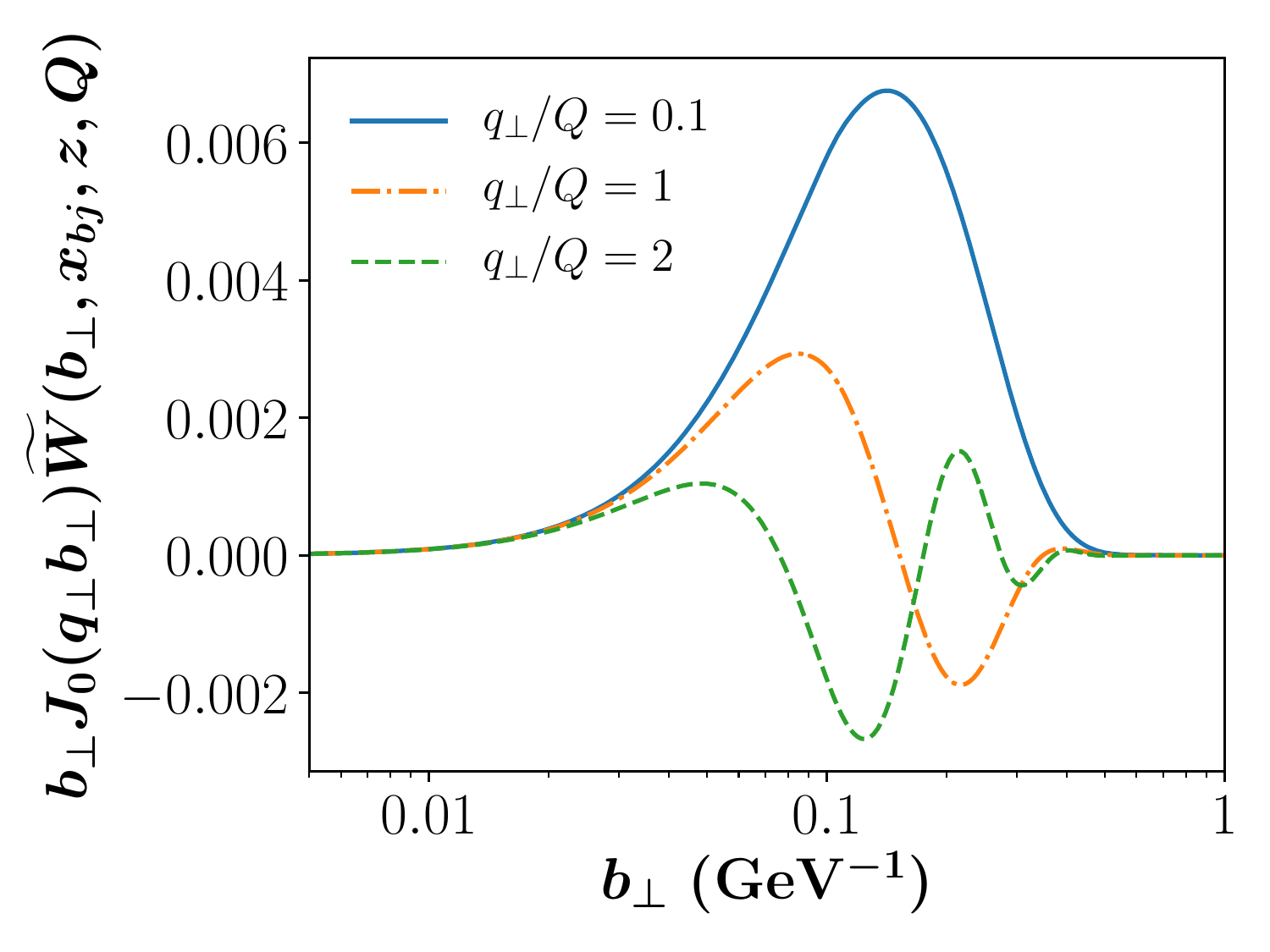}
\caption{
  SIDIS $\bt$ space integrand from \eref{Fuu} for the differential
  cross section at NLO+NLL for $\pi^+$ production for three different
  values of transverse momenta: $\qt/Q = 0.1$, $1$, and $2$,
  respectively.  For the rest of the external kinematics we select
  $S_{\ell p} = 52.7$ (GeV$^2$), $\xbj = 0.25$, $z = 0.5$, and $Q^2 =
  2.5$ (GeV$^2$) within the coverage of the HERMES experiment~\cite{Airapetian:2012ki}.}
\label{f.TMDbT}
\end{figure}
The expression for the $W$ term in~\eref{Fuu} is given by
\begin{align}
    \widetilde{W}(\bt, \xbj, z , Q) = & H(Q,\mu)\sum_q e_q^2 \; C_{q/j}^{pdf}\otimes f_{j/p}(\xbj, \mu_{b*}) \; C_{i/q}^{ff}\otimes d_{h/i}(z, \mu_{b*}) \nonumber\\ & \times\exp\left[2S_{\rm pert}-(g_q+g_h/z^2)\;\bt^2-g_2\ln\left(\frac{\bt}{b_{*}}\right)\ln\left(\frac{Q}{Q_0}\right)\right] \;
\end{align}
where $\otimes$ is the convolution operator given in Eqs.~(38) and (39) in \cite{Kang:2015msa}.

Having established the QCD based TMD setups, let's examine the
behavior of the SIDIS cross section in $\bt$ space.  In \fref{TMDbT}
the $\bt$ space integrand given in \eref{Fuu} for the SIDIS
differential cross section is plotted, for three different values of
$q_\perp/Q = 0.1,~1.0$ and $2.0$, respectively. We take the
lepton-proton center-of-mass energy square $S_{\ell p} = 52.7
$~(GeV$^2$), $\xbj = 0.25$, $z = 0.5$, and $Q^2 = 2.5$ (GeV$^2$).
These kinematics are within the coverage of the pion production in
unpolarized lepton-proton SIDIS data at the HERMES experiment
~\cite{Airapetian:2012ki}.  Just like in the case of the toy
TMDs in \sref{Benchmarking the Numerical Precision}, the integrand
becomes more oscillating as $\qt/Q$ increases. As a consequence, the
numerical estimation of the  Fourier-Bessel transform from $\bt$-space
to  $\qt$-space  becomes increasingly more challenging for larger
values of $q_{\perp}/Q$. 

\begin{figure}[htb!]
\centering
\includegraphics[width=\textwidth]{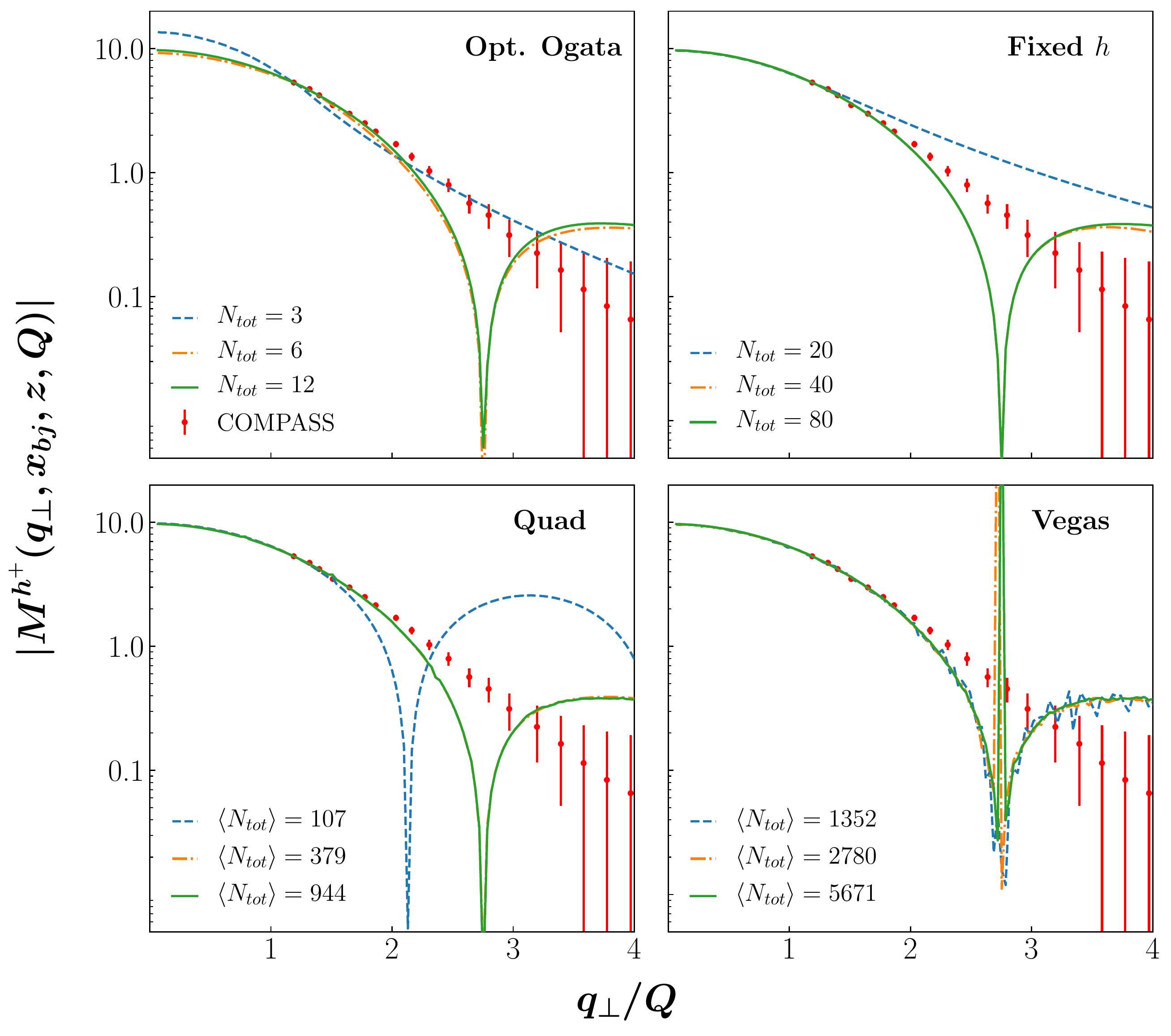}
\caption{
  The theoretical prediction for the hadron multiplicity,
  $|M^{h^+}(\qt,\xbj,z,Q)|$, as a function of $\qt/Q$, using four different
  integration algorithms: the optimized Ogata quadrature (``Opt. Ogata'',
  top left panel), the standard Ogata method (``Fixed $h$ Ogata'', top right panel), the adaptive Gaussian quadrature (``Quad'', bottom left panel), and the
  Vegas Monte Carlo algorithm (``Vegas'', bottom right panel). For illustration, we
  choose the kinematics to be consistent with the COMPASS experiment:
  $0.02<\xbj<0.032$, $z= 0.2$, and 1.7 GeV$^2$ $< Q^2 <$ 3 GeV$^2$. The
  experimental data from COMPASS~\cite{Aghasyan:2017ctw} are also shown
  for comparison (red solid points). 
}
\label{f.Three-Three}
\end{figure}
We next analyze the performance of our optimized Ogata
quadrature to get the $\qt$-space cross sections against the fixed $h$ Ogata, adaptive
Gaussian quadrature, and the Vegas Monte Carlo integration routines.
For that we consider a related experimental observable, the
hadron multiplicity which has been measured by both
HERMES~\cite{Airapetian:2012ki} and
COMPASS~\cite{Aghasyan:2017ctw} experiments. The COMPASS hadron multiplicity
is defined as~\cite{Aghasyan:2017ctw}
\begin{align}
M^h(\qt,\xbj,z,Q) = \frac{\pi}{z^2}\left. 
\frac{d\sigma^h}{d\xbj dydzd^2\qt}\right/ \frac{d\sigma^{DIS}}{d\xbj dy}\; ,
\label{e.Mh}
\end{align}
where the numerator is the SIDIS cross section for the production of a
hadron $h$ that we have been discussing so far, and the denominator is
the inclusive DIS cross section.
In \fref{Three-Three}, we plot the
absolute value of the theoretical prediction for the hadron
multiplicity, $|M^{h^+}(\qt,\xbj,z,Q)|$, as a function of $\qt/Q$, using
the above mentioned integration algorithms.
For illustration, we choose the kinematics to be
consistent with hadron multiplicity data from COMPASS experiment:
$0.02<\xbj<0.032$, $z= 0.2$, and 1.7 GeV$^2$ $< Q^2 <$ 3 GeV$^2$. The
four panels correspond to optimized Ogata in the top left
(labeled as `Opt. Ogata'), standard Ogata in the top right 
(labeled as `Fixed $h$ Ogata'), adaptive Gaussian quadrature in the 
bottom left (labeled as Quad), and Vegas Monte Carlo in the bottom 
right (labeled as Vegas), respectively. We also plot the COMPASS experimental data in \fref{Three-Three} for
comparison.~\footnote{In order to describe the data, the normalization
issue with the COMPASS data must be resolved. We follow the work done
in \cite{Bacchetta:2017gcc} to normalize the COMPASS multiplicities
such that the data and theory are equal at the lowest values of the
transverse momentum in each $z$ bin.} 

Note that at relatively large hadron transverse momentum
$\qt/Q\gtrsim2$, the theoretical calculations in TMD factorization
formalism would become negative. There, one has to include the so-called
$Y$-term~\cite{Collins:2011zzd}, or switch/match onto the usual
collinear factorization formalism~\cite{Collins:1984kg,Landry:2002ix}.
For very large values of $Q$, such an switching/matching from TMD factorization to 
simply a collinear factorization is straightforward and usually happens when $\qt$ becomes very large $\qt\sim Q$.
It turns out that for small values of $Q$ (order of several GeV) which are most relevant for data at HERMES, COMPASS, and JLab, the matching is very tricky and does not occur as a sharp transition in the usual $W+Y$ prescription. 
Due to this, it is very important that in phenomenological applications the matching can be implemented without large numerical errors stemming from the Bessel transform using an efficient algorithm such as the one we are proposing. In the vicinity of $q_\perp \sim Q$ the calculation should be specifically precise in order to allow for the transition to happen. 
Moreover since the precise location of where the transition occurs can only be confronted phenomenologically, it is important that the ability to transform the $W$ term into $q_{\perp}$ space is as precise as possible to avoid any biases for the TMD extraction.
See,  e.g.~Refs.~\cite{Collins:2016hqq,Boglione:2014oea} for more details. It is because of this reason that our demonstration in \fref{Three-Three} covers the broad region of $\qt$, from the small $\qt\ll Q$ to much larger $\qt\gtrsim Q$.  

For each integration method in this \fref{Three-Three}, the number of nodes is increased until the relative error of the inversion associated with doubling the number of nodes is smaller than the total experimental uncertainty for all data points in the set~\footnote{In other applications of our optimized Ogata quadrature algorithm, the same method can be used to estimate the values for $N$ for a given integrand and desired precision.}. This curve is plotted in orange, while in green and blue we plot the inversions with roughly half and double the number of nodes as the orange curve. For the optimized and fixed $h$ Ogata methods, we provide the total number of calls to the integrand $N_{tot}$. 
We note that the total number of calls to the integrand for the optimized Ogata contains two contributions.
The algorithm first uses the gradient of the $h$-space function $h(h\xi_{n1})^{2n+1}\widetilde{W}(\xbj,z,h\xi_{n1}/\qt,Q)$ to determine $h_{opt}$ in Eqs.~\eqref{e.opt} and \eqref{e.opt_t}.
The numerical algorithm then samples the nodes to perform the quadrature sum. 
The first contribution tends to require more sampling points than the second contribution.
However,  the perturbative factors as well as the collinear distribution functions in this expression  vary much more slowly in $h$-space than the non-perturbative factors. We find that this tends to be true in general for processes with small $Q$, where non-perturbative TMD structure is more sensitive.   
One can then sample only the expression $h(h\xi_{n1})^{2n+1}(h\xi_{n1}/\qt e^{-S_{NP}})$ to determine the value of $h_{opt}$.
Since sampling this expression requires little computational power, we do not count  this contribution to $N_{tot}$. 
We also note that in principle in an actual fit where a $\chi^2$ minimization is performed, one does not need to estimate the $h$ values for every step of $\chi^2$ evaluation. Instead these values can be updated every certain number of steps during the minimization. This will further enhance the efficiency of our method compared to the other approaches. In the adaptive quadrature and Vegas Monte Carlo methods, we indicate the average number of calls to the integrand $\left<N_{tot}\right>$. 

As one can note, in the limit of large sampling, all the numerical integrators converge to the same result. However, the optimized Ogata quadrature converges to this result nearly an order of magnitude faster than the fixed $h$ method, nearly two orders of magnitude faster than adaptive Gaussian quadrature, and nearly three orders of magnitude faster than Vegas Monte Carlo integration. This result demonstrates that our optimized Ogata algorithm can improve significantly the numerical efficiency of the Fourier-Bessel integration encountered in the TMD analysis.
\begin{figure}[t!]
\centering
\includegraphics[width=0.75\textwidth]{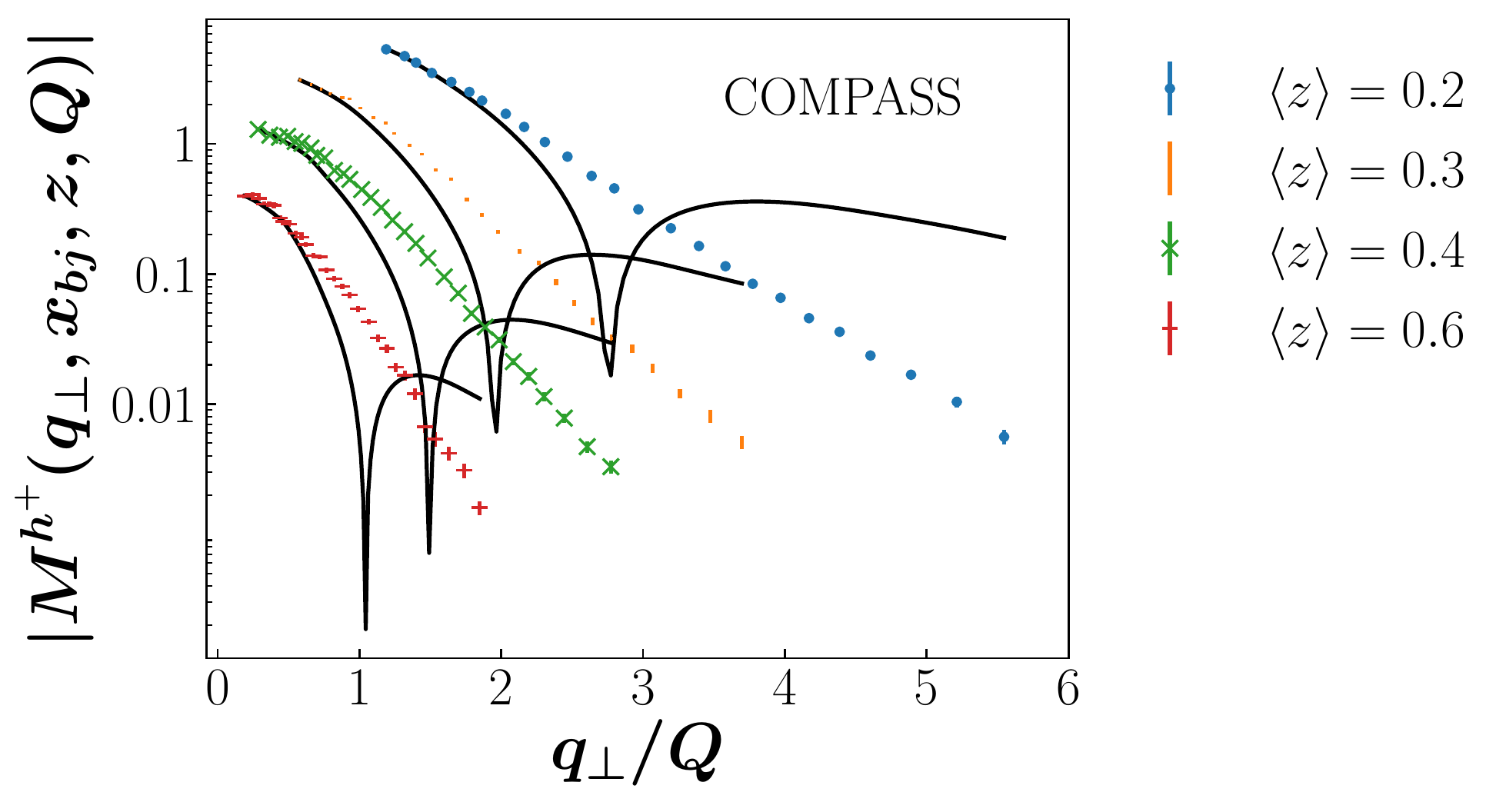}
\caption{Comparison of COMPASS hadron multiplicity data
  \cite{Aghasyan:2017ctw} and absolute value of the theoretical calculations using the optimized Ogata algorithm with a number of nodes $N = 6$. $N = 6$ was chosen as it gave reliable results over the entire region in \fref{Three-Three}. The computations are performed for $0.02<\xbj<0.032$, 1.7 GeV$^2$ $< Q^2 <$ 3 GeV$^2$, and for 4 different $\langle z\rangle$ values as shown in the figure. The black solid curves are the absolute values of the theoretical results.}
\label{f.COMPASS}
\end{figure}

Finally in \fref{COMPASS}, we plot four multiplicity distributions at
different values of $\langle z\rangle = 0.2,0.3,0.4,0.6$, respectively
for the bins $0.02<\xbj<0.032$ and 1.7 GeV$^2$ $<Q^2<$3 GeV$^2$ using
the optimized Ogata algorithm with the number of nodes $N
= 6$. $N = 6$ was chosen as it gave reliable results over the entire region in \fref{Three-Three}. It is worthwhile to emphasize again that the theory predictions~\footnote{Note that the $q_\perp$-distribution in the TMD formalism gets narrower as $\langle z\rangle$ increases. This is expected from theoretical consideration, see e.g. Ref.~\cite{Anselmino:2013lza}.}
become extremely efficient, thanks to the optimized Ogata quadrature.
This gives us a great confidence that the optimized Ogata method would
be ideal in the future for performing efficient numerical calculations and/or for the global analysis of TMDs.~\footnote{Note that we are not presenting a new fit here. Rather we just display, using fixed parameters from
Refs.~\cite{Kang:2015msa,Su:2014wpa}, that this numerical method can be used to perform efficient numerical calculations for describing TMD data.}
\section{Conclusions}
\label{s.conclusions}

In this paper we have developed a high performance numerical algorithm
for Hankel transforms for TMD factorization formalism from position
$\bt$ space to transverse momentum $\qt$ space using the optimized
Ogata quadrature method, which uses the zeros of Bessel functions as
nodes.  For a relatively small and fixed number $N$ of functional
calls to the integrand, we derived conditions to find the optimal
parameter $h$, which controls the node density. Such an optimized
Ogata quadrature ensures a 
small number of calls while achieving a
high accuracy at the same time, and thus becomes extremely efficient
in TMD studies. We use both toy TMDs, and parametrizations of QCD
based TMDs to demonstrate the efficiency of our integration algorithm.
We found that the optimized Ogata quadrature performs nearly an order of magnitude faster than standard Ogata methods, nearly two orders of magnitude faster than adaptive Gaussian quadrature, and nearly three orders of magnitude faster than Vegas Monte Carlo integration for all regions of transverse momentum in semi-inclusive deep inelastic scattering. Our algorithm thus can have wide application in the future TMD computations and/or TMD global analysis. The code which illustrates the optimized Ogata quadrature is available for download in Python2, Python3, C++ with Boost dependency, C++ with GSL dependency, and a standalone Fortran 77 with an open source license at \href{https://github.com/UCLA-TMD/Ogata}{https://github.com/UCLA-TMD/Ogata}. For information on installation and usage visit
\href{https://ucla-tmd.github.io/Ogata/}{https://ucla-tmd.github.io/Ogata/}.

\section*{Acknowledgements}
Z.K. is supported by the National Science Foundation under Grant No.~PHY-1720486 and No.~PHY-1945471.  A.P. is supported by the National Science Foundation under Grant No.~PHY-1623454 and the DOE Contract No. DE-AC05-06OR23177, under which Jefferson Science Associates, LLC operates Jefferson Lab. N.S. was supported by the DOE contract DE-SC0018106. J.T. is supported by the National Science Foundation under Grant No.~DGE-1650604. This work is also supported within the framework of the TMD Topical Collaboration.
\bibliography{refs}
\bibliographystyle{h-physrev5}

\end{document}